\documentclass[journal,twoside,web]{ieeecolor}
\usepackage{generic}
\usepackage{cite}
\usepackage{amsmath,amssymb,amsfonts}
\usepackage{algorithmic}
\usepackage{graphicx}
\usepackage{textcomp}
\usepackage{multirow}
\usepackage{multicol}
\def\BibTeX{{\rm B\kern-.05em{\sc i\kern-.025em b}\kern-.08em
    T\kern-.1667em\lower.7ex\hbox{E}\kern-.125emX}}
\begin{document}
\title{Myocardial Segmentation of Cardiac MRI Sequences with Temporal Consistency for Coronary Artery Disease Diagnosis}
\author{Yutian Chen, Xiaowei Xu, Dewen Zeng, Yiyu Shi, Haiyun Yuan, Jian Zhuang, Yuhao Dong, Qianjun Jia, Meiping Huang

\thanks{This work was supported by the Science and Technology Planning Project of Guangdong Province under Grant No. 2017A070701013, 2017B090904034, 2017B030314109, 2018B090944002, and 2019B020230003, Guangdong peak project under Grant No. DFJH201802, the National Key Research and Development Program under Grant No. 2018YFC1002600, the Natural Science Foundation of Guangdong Province under Grant No. 2018A030313785.}
\thanks{Yutian Chen, Xiaowe Xu and Dewen Zeng contribute equally to this work. Yuhao Dong, Qianjun Jia and Meiping Huang contribute equally to this work.}
\thanks{Yutian Chen and Xiaowe Xu are with Guangdong Cardiovascular Institute, and Haiyun Yuan and Jian Zhuang are with the Department of Cardiovascular Surgery, and Yuhao Dong, Qianjun Jia, and Meiping Huang are with the Department of Catheterization Lab. They are all with Guangdong Provincial Key Laboratory of South China Structural Heart Disease, Guangdong Provincial People's Hospital, Guangdong Academy of Medical Sciences, 106 Zhongshan Second Road, Guangzhou, China (e-mail: xiao.wei.xu@foxmail.com). }
\thanks{Dewen Zeng and Yiyu Shi are with Department of Computer Science and Engineering University of Notre Dame, IN, US, 46556 (e-mail: yshi4@nd.edu).}}

\maketitle

\begin{abstract}
Coronary artery disease (CAD) is the most common cause of death globally, and its diagnosis is usually based on manual myocardial segmentation of Magnetic Resonance Imaging (MRI) sequences.
As the manual segmentation is tedious, time-consuming and with low replicability, automatic myocardial segmentation using machine learning techniques has been widely explored recently.
However, almost all the existing methods treat the input MRI sequences independently, which fails to capture the temporal information between sequences, e.g., the shape and location information of the myocardium in sequences along time.
In this paper, we propose a myocardial segmentation framework for sequence of cardiac MRI (CMR) scanning images of left ventricular cavity, right ventricular cavity, and myocardium.
Specifically, we propose to combine conventional networks and recurrent networks to incorporate temporal information between sequences to ensure temporal consistent. We evaluated our framework on the Automated Cardiac Diagnosis Challenge (ACDC) dataset. Experiment results demonstrate that our framework can improve the segmentation accuracy by up to 2\% in Dice coefficient.
\end{abstract}


\section{Introduction}

Coronary artery disease (CAD) is the most common cause of death globally.
It affects more than 100 millions of people, and results in about 10 millions death each year \cite{vos2016global}.
In the United States, about 20\% of those over 65 have CAD \cite{centers2011prevalence}.
Magnetic resonance imaging (MRI) is a common tool for CAD diagnosis. With cardiac MRI (CMR), the myocardial structure and functionality can be assessed and analyzed.
Particularly, experienced radiologists manually perform myocardial segmentation on the CMR image sequences and measures several parameters to finally determine the diagnosis. For instance, the left and right ejection fractions (EF) and stroke volumes (SV) are widely used for cardiac function analysis \cite{8360453}.

Recently, automatic myocardial segmentation of CMR image sequences has attracted considerable attention in the community.
On one hand, with the aging society, the number of patients with CAD has been increasing for decades \cite{odden2011impact}. 
On the other hand, manual myocardial segmentation is tedious, time-consuming and with low replicability. 
Considering the medical cost and quality, automatic myocardial segmentation is highly desirable.
However, it is a challenging task.
First, there exists large shape variations in the images.
Second, the labels of the noisy images is with low uniformity which degrades the training efficiency and effectiveness.

Currently, there exists two approaches for automatic myocardial segmentation.
In the traditional myocardial segmentation approach \cite{li2016myocardial,guo2017novel}, an manually defined contour or boundary is needed for initialization. 
Although an automatic initialization might be achieved by some algorithms \cite{barbosa2013fast,van2008time}, the segmentation performance highly relies on the initialization quality, which makes the framework lack stability. 
Another approach \cite{leclerc2017fully, chen2016iterative} uses deep learning for myocardial segmentation, which do not need any initialization and the whole process can run without manual interaction.
However, these methods treat each CMR frame independently which do not exploit the temporal consistence among sequences.

In this paper, we propose to exploit temporal consistence for myocardial segmentation of CMR sequences for automatic CAD diagnosis.
Particularly, we propose an encoder-decoder framework combining conventional networks and recurrent neural networks.
The encoder is able to extract a set of features from CMR sequences, while the decoder embeds convolutional networks into recurrent networks to incorporate temporal information between CMR sequences.
The contributions of our work are:

\begin{itemize}

\item We proposed an encoder-decoder framework for myocardial segmentation of CMR sequences. our framework is able to incorporate temporal information between CMR frames. 

\item To further exploit the temporal consistence among frames, we adopted a bi-directional training approach which can reduce segmentation error introduced by the first few frames in the training process. 

\item We conducted comprehensive experiments on the ACDC dataset. Compared with the residual 3D U-net model of \cite{Yang2018}, our framework achieves an improvement of 1\%-2\% of segmentation accuracy in Dice coefficient.

\end{itemize}

\section{Background}

\subsection{Myocardial Segmentation of CMR Image}

CMR image is a widely used imaging tool for the assessment of myocardial micro-circulation. It utilizes the electromagnetic signal with characteristic frequency produced by the hydrogen nuclei under strong contrasting magnetic field and weak oscillating near field as the imaging agent.

Due to the high capacity for discriminating different types of tissues, CMR image is one of the most prominent standards for cardiac diagnosis through the assessment of the left and right ventricular ejection functions (EF) and stroke volumes (SV), the left ventricle mass and the myocardium thickness. For example, \cite{8360453} obtained these parameters from CMR images using an accurate segmentation of CMR image for left ventricular cavity, right ventricular cavity, and the myocardium at end diastolic (ED) frame and end systolic (ES) frame can give out an accurate diagnostic of cardiac function. 

\begin{figure}[h]
  \centering
  \includegraphics[width=1.0\columnwidth]{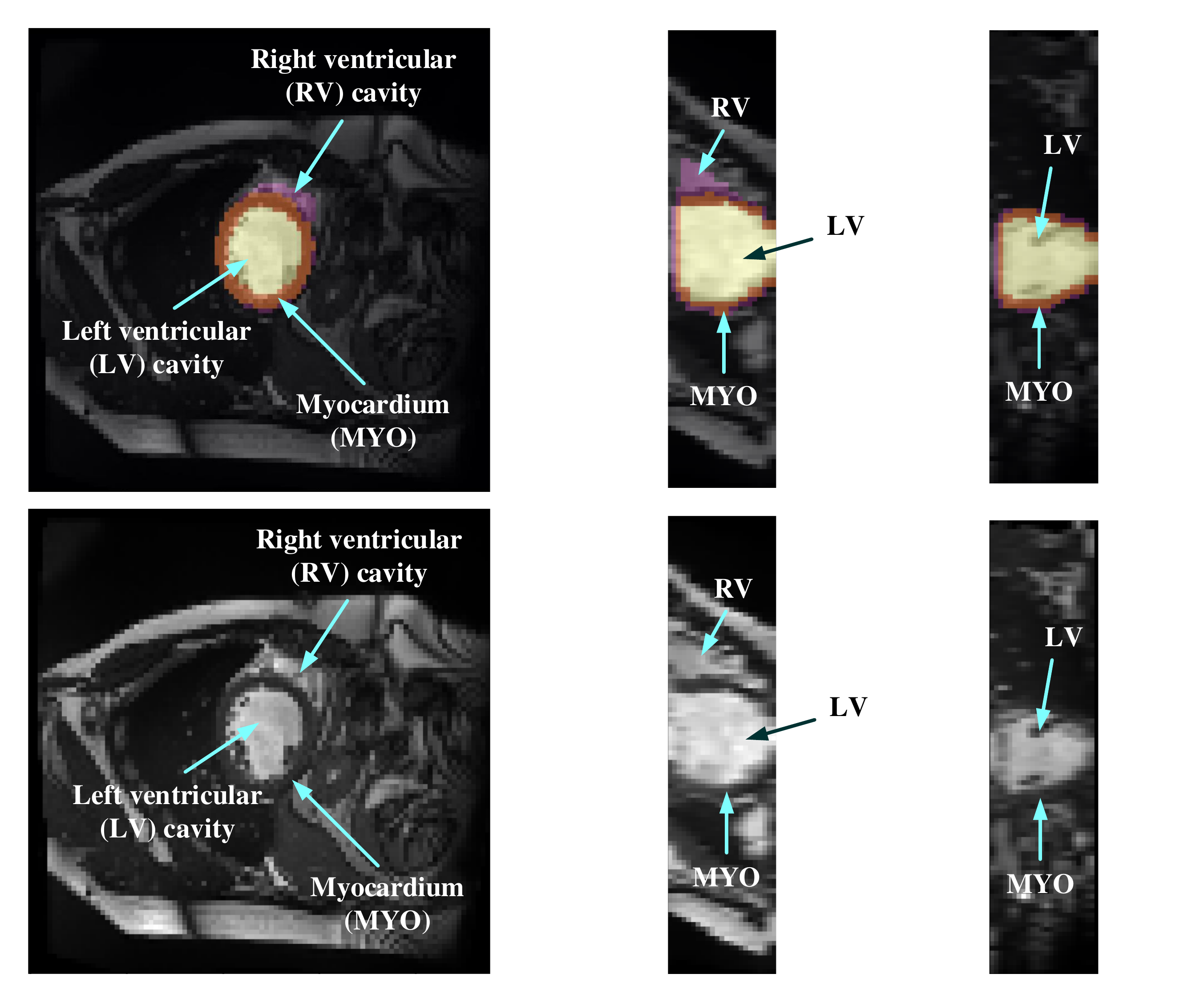}
  \caption{Structure illustration of a typical CMR image. Images above are the slices on z-axis, y-axis and x-axis respectively from patient 001 in the ACDC dataset at ED frame with mask, and the second row of figure are the raw CMR slices of patient 001.}
  \label{fig1}
\end{figure}

In order to evaluate the myocardial function, accurate segmentation of left ventricular (LV) cavity, right ventricular (RV) cavity, and myocardial (MYO) need to be acquired from the framework. Figure \ref{fig1} shows the slices of typical CMR images of a patient at ED frame with and without ground truth mask along the each axis respectively. The label shows the ground truth of segment result for different parts of the CMR image.

\subsection{Related Work}
Myocardial segmentation of CMR sequences has the following challenges.
First, the contrast between myocardium and surrounding structures are low as shown in Figure \ref{fig1}.
Second, the brightness heterogeneity in the left and right ventricular cavities due to blood flow \cite{8360453}.
Third, misleading structures such as papillary muscle have the same intensity and grayscale information as myocardium, which makes it hard to extract the accurate boundary. 
There are two approaches among existing works towards myocardium segmentation.

The first approach is based on point distribution models (PDMs) \cite{tobon2008automatic}.
A good example is the active shape model (ASM) \cite{cootes1995active} or active appearance model (AAM) \cite{cootes2001active}.
The main idea of ASM is to learn patterns of variability from a training set of correctly annotated image.
ASM uses principal component analysis (PCA) to build a statistical shape model from a set of training shapes, and then fits an image in a way that most similar to the statistical shape in the training set.
\cite{inproceedings} proposed an algorithm for myocardial and left ventricular cavity segmentation in CMR images based on invariant optimal feature 3-D ASM (IOF-ASM).
\cite{vanAssen2006} improved the ASM such that the method can work for sparse and arbitrary oriented CMR images.
\cite{Tobon_Gomez_2012} proposed a new ASM model that includes the measurement of reliability during the matching process to increase the robustness of the model.
\cite{Santiago2016} proposed a method of applying ASM on CMR images with varying number of slices and perform segmentation on arbitrary slice of CMR images with a new re-sampling strategy. 

The prediction results of ASM must be constrain into certain shape variations so that the shape of the segmentation result does not go too far from the regular myocardium shape.
Note that this is very important when artifacts and defects in the CMR image make the myocardium boundary unclear and hard to recognize. 
However, ASM is based on linear intensity information in the image, which is insufficient to model the appearance of CMR data with huge intensity variations and large artifacts. 
In addition, ASM requires a manual initialization shape and the final segmentation result is very sensitive to the shape and position of this initialization. 
Thus a fully automatic and non-linear model is needed.

The second approach adopts machine learning techniques to perform image segmentation.
For example, \cite{Zhang2016} used a simple implementation of fully connected neural network for the quality assessment of CMR image.
\cite{poudel2016recurrent} used recurrent fully connect network (RFCN) on stack of 2D image for segmentation of CMR images. The recurrent network is applied on the short axis so that the continuous of spacial information on the short axis can be utilized. 
\cite{9153905} proposed to use Dilation CNN, where each layer has the same resolution so the localized information in the input image would not be lost.
\cite{Isensee2018} proposed a multi-structure segmentation for each time step of MRI sequences and extracted a domain-specific features.
\cite{Simantiris2020} used a simple network composed of cascaded modules of dilated convolutions with increasing dilation rate without using concatenation or operations like pooling that will lead to the decrease of resolution.
\cite{Zotti2019} introduced the shape prior obtained from the training dataset in the 3D Grid-net and employed the contour loss as loss function to improve the performance on the border of segmentation result.
\cite{Painchaud2019} presented a method to guarantee the anatomical plausibility of segmentation result such that the anatomical invalid segmentation result of the model will be reduced to zero.
\cite{Khened2018} proposed a neural network with the dense block that contains dense connections between layers inspired by the Dense Net.
\cite{Baumgartner2018} compared the performance of 2D and 3D fully convolution network (FCN) and U-net.
\cite{BaldeonCalisto2020} used an multi-objective evolutionary based algorithm to incorporate 2D FCN and 3D FCN to search for an efficient and high-performing architecture automatically.
\cite{Wolterink2018} used six different types of model's average probability map and cyclic learning rate schedule to improve the segmentation performance.
\cite{Roh2018} proposed a combination of rigid alignment, non-rigid diffeomorphism registration, and label fusion to increase the performance of 3D U-net.
\cite{Zotti2018} used the shape prior that is embedded in the GridNet to reduce the anatomical impossible segmentation result.
\cite{Patravali2018} used the combination of 2D and 3D U-net and proposed a new class-balanced Dice loss to make the optimization easier.

Although these above methods showed great improvements in the segmentation performance compared to ASM or AAM. 
They treat each frame independently, which make the segmentation results of some specific sequences inaccurate or the overall results lack coherence.

\subsection{Dataset}

\begin{figure}[]
  \centering  
  \includegraphics[width=1\columnwidth]{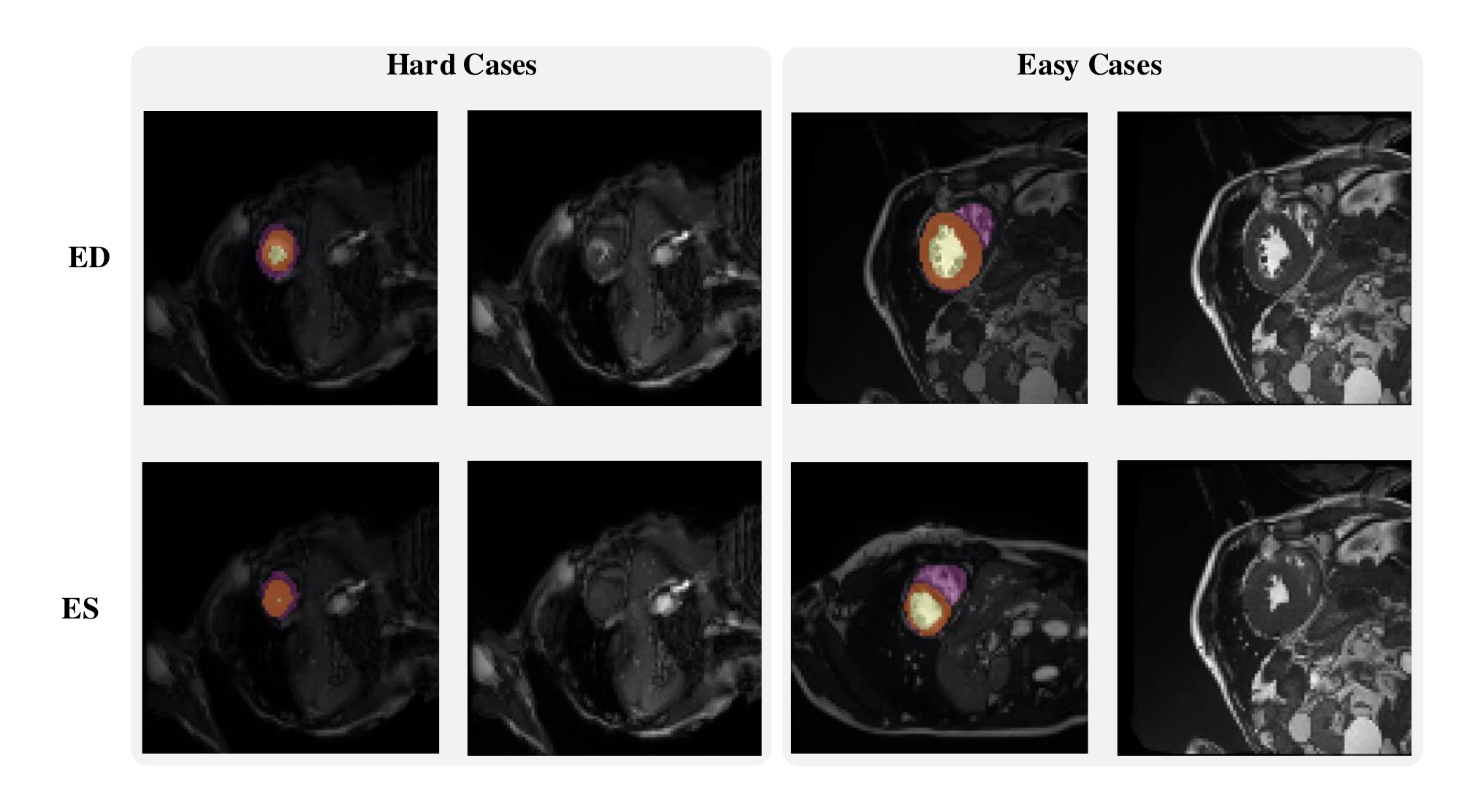}
  \caption{Examples of hard and easy cases of CMR image (slice taken on short-axis). The first and second columns refer to hard cases and the third and forth columns refer to easy cases.}
  \label{fig2}
\end{figure}

The automatic cardiac diagnostic challenge (ACDC) dataset consists of both CAD patients and healthy individuals, whose diagnosis results are extracted from clinical medical cases. 
There are 150 patients in total and are evenly divided into five subgroups base on their diagnosis results. The five subgroups of patients have systolic heart failure with infraction, dilated cardiomyopathy, hypertrophic cardiomyopathy, abnormal right ventricle, and no abnormality, respectively. 50 of the patients made up the test dataset on the ACDC website, and the other patients are released as the training dataset.
CMR sequences of all patients are collected by two MRI systems with different magnetic strength (1.5T-Siemens Area, Siemens Medical Solutions, Germany and 3.0T-Siemens Trio Tim, Siemens Medical Solutions, Germany). For each frame in the patient's CMR sequence, there contains a series of short-axis slices covering the LV from base to apex \cite{8360453}.
For most patients, the dataset collected 28 - 40 consecutive frames to cover the whole cardiac cycle. Some of the patients in the dataset may have 5-10 percents of the cardiac cycle being omitted.

Figure \ref{fig2} shows some hard and easy cases in both ED and ES phases of CMR image in the ACDC dataset. The hard cases usually have a low contrast, blur image, or extreme anatomical structure. While the easy cases have a high contrast, and less misleading structure with the similar features as LV, RV, and MYO have.

\section{Method}

\begin{figure}[]
  \centering
  \includegraphics[width=1\columnwidth]{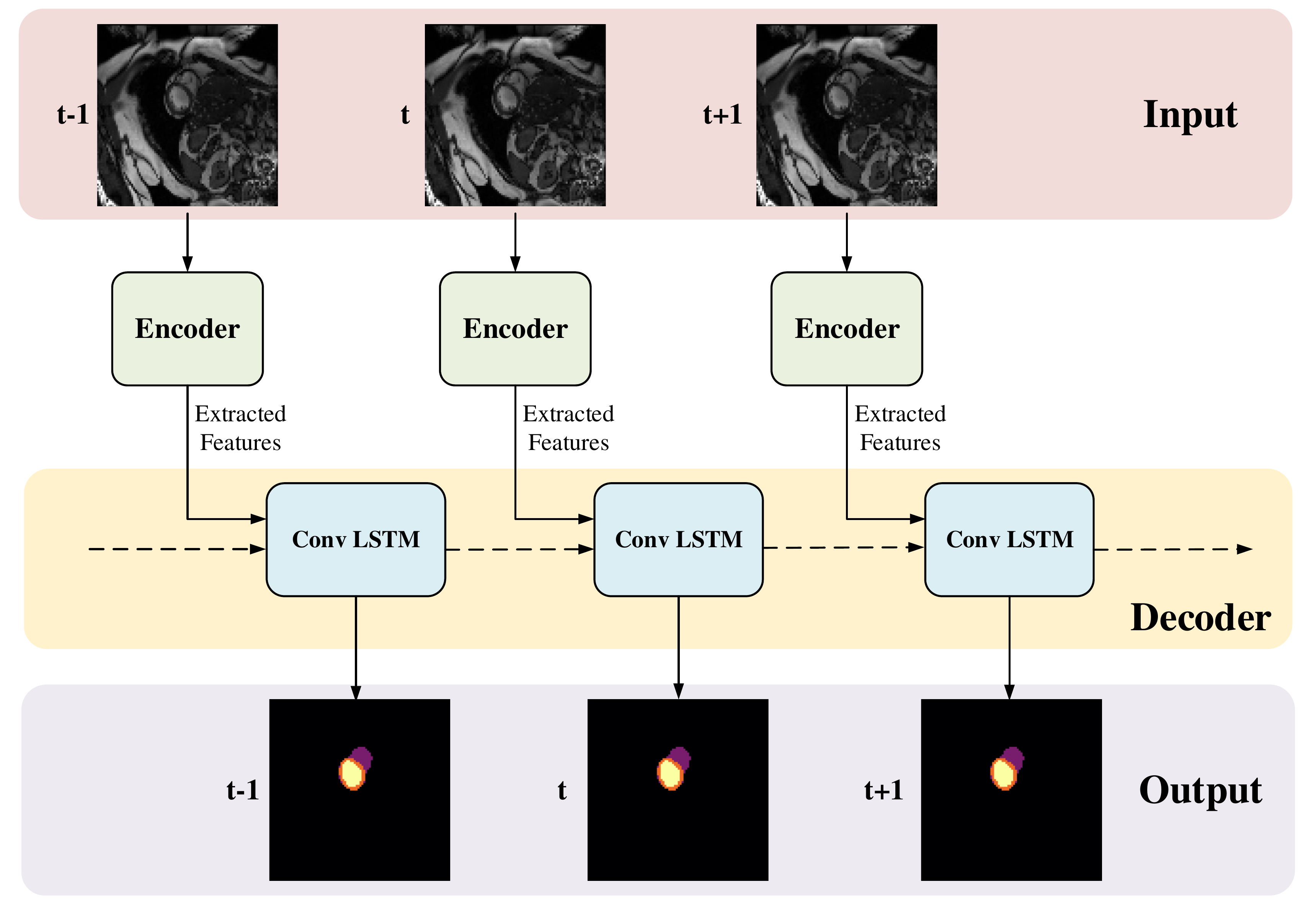}
  \caption{The proposed myocardial segmentation architecture, which contains an encoder (Res U-Net \cite{ronneberger2015u}) and a decoder (ConvLSTM \cite{salvador2017recurrent}).}
  \label{fig3}
\end{figure}

\begin{figure}[]
  \centering
  \includegraphics[width=1\columnwidth]{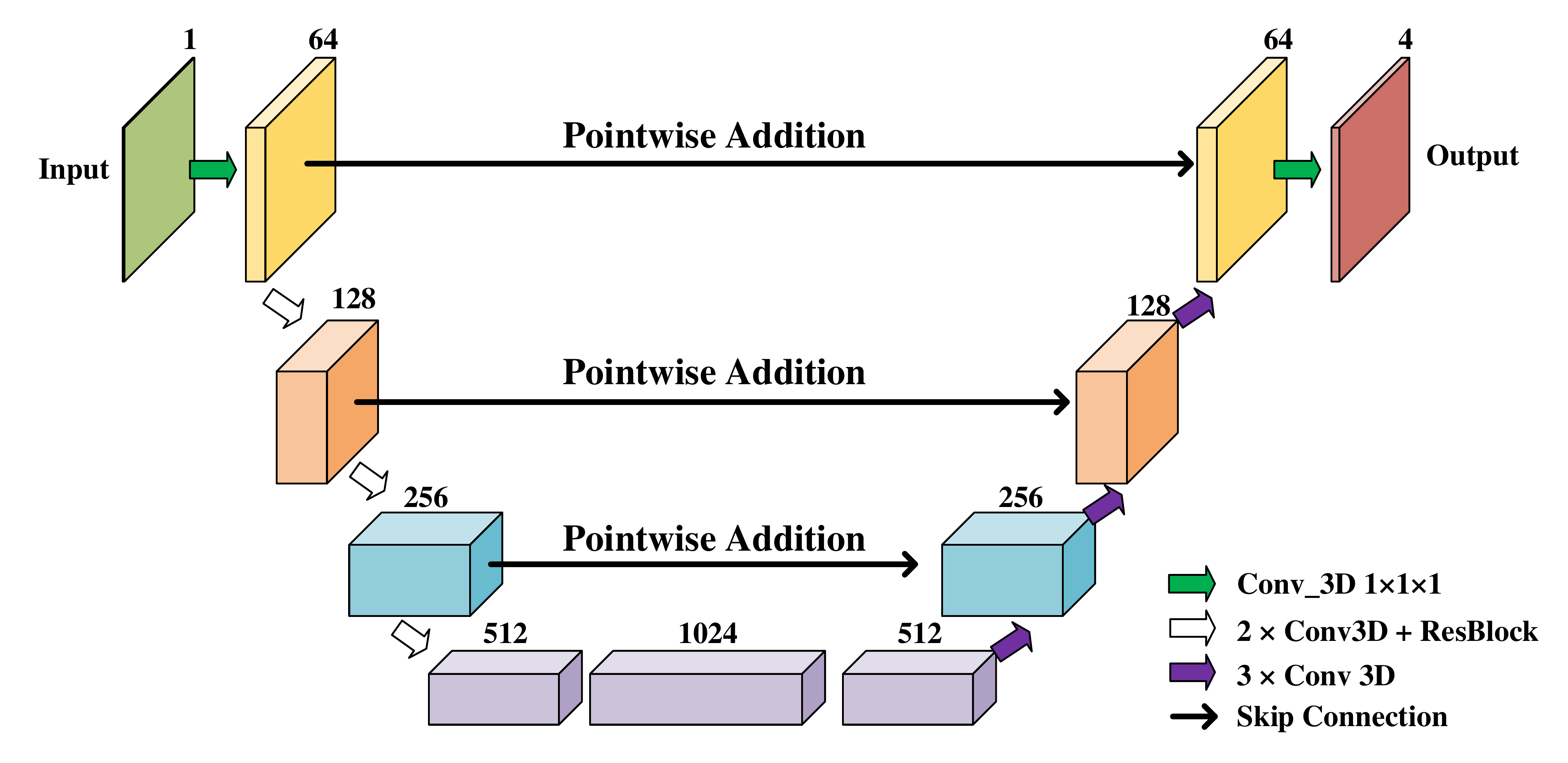}
  \caption{Network structure of our proposed Res U-net based encoder.}
  \label{fig4}
\end{figure}

\begin{figure*}[]
  \centering
  \includegraphics[width=1.9 \columnwidth]{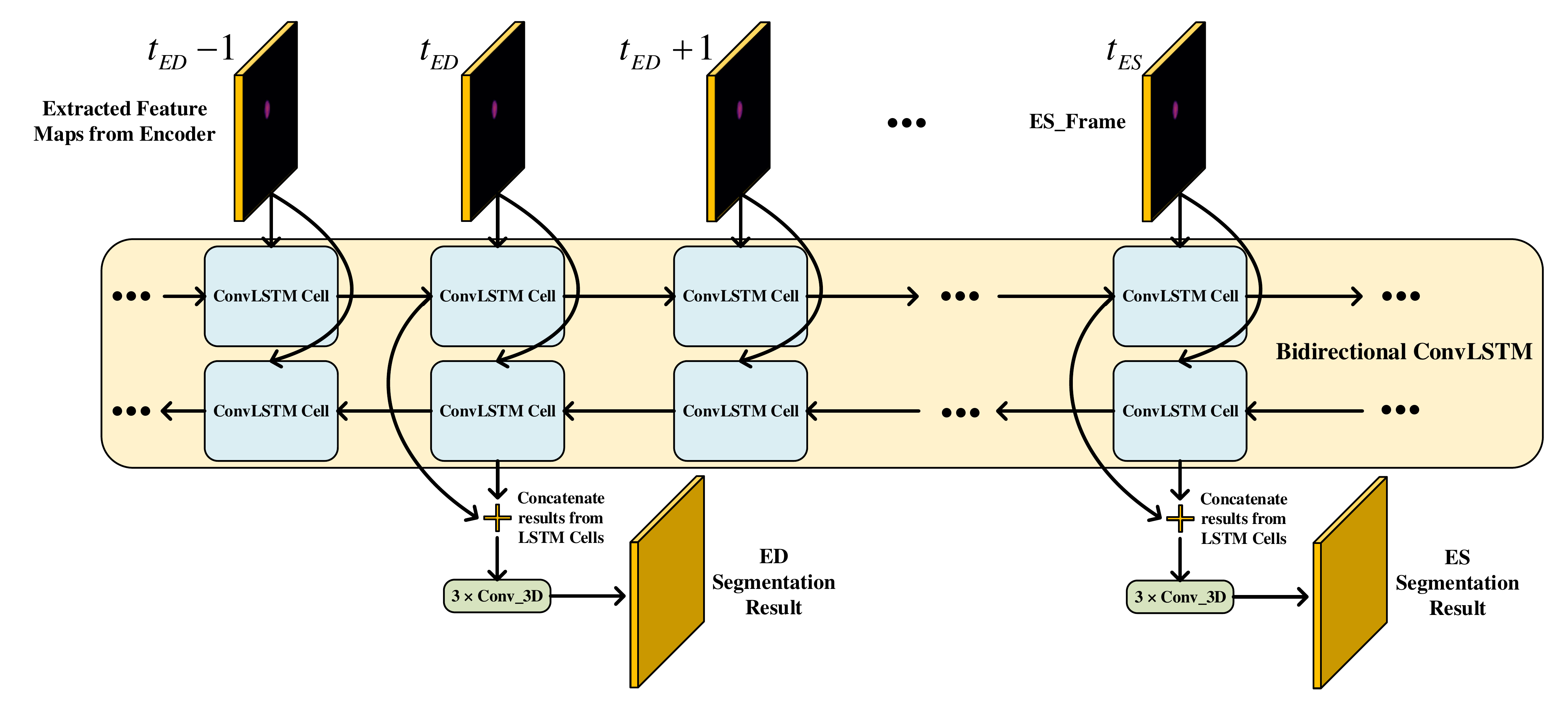}
  \caption{Illustration of our proposed bi-direction training approach.}
  \label{fig7}
\end{figure*}

The proposed framework for myocardial segmentation of CMR image sequences is shown in Figure \ref{fig3}. The input consists of a set of CMR frames from a CMR sequence. The output is the myocardial segmentation at ED and ES phases of the input.

\begin{figure*}[]
  \centering
  \includegraphics[width=1.9\columnwidth]{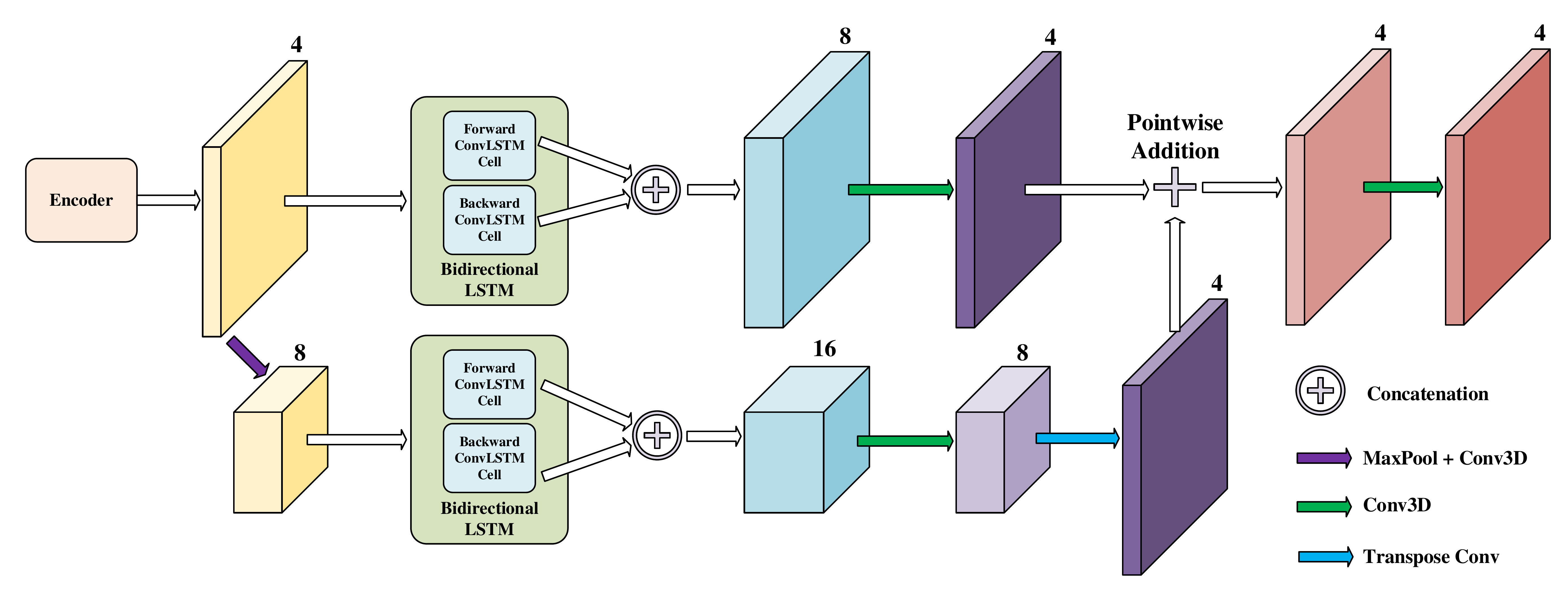}
  \caption{Network structure of our proposed decoder. The input of our decoder is the features extracted from our encoder. The decoder consists of a hierarchical ConvLSTMs and is able to incorporate temporal information between CMR frames.}
  \label{fig5}
\end{figure*}

\subsection{Encoder}

The encoder is based on U-net \cite{ronneberger2015u}, which is an effective method on a broad range of medical image segmentation tasks. 
The network structure of the encoder is shown in Figure \ref{fig4}. 
The input of our encoder is a single-channel image corresponding to one frame in a MCE sequence.
Based on the U-net, we add one residual block on each layer, which contains three convolutional layers and one shortcut path in it. Also, the results from each layer are combined using pointwise addition instead of concatenation. We expect the addition of residual blocks in U-net can 
extract features from the input CMR image without suffering from serious gradient explosion or gradient vanishing problem.

We extract four feature maps from U-net as the output of our encoder. 
Four feature maps obtained from the encoder represents the probability that one voxel belongs to background, LV, RV, and MYO, respectively.

\subsection{Decoder}
The network structure of the decoder is shown in Figure \ref{fig5}. It contains a hierarchical recurrent network of ConvLSTMs \cite{xingjian2015convolutional} which acts like recurrent U-net. 
The output of the decoder is the segmentation result of myocardium of this frame.
The dash arrows in Figure \ref{fig3} depict the temporal recurrence in the decoder.
We depict the features extracted by the encoder for frame $t$ as $f_{t}$, and $g_{t, k}$ as the feature for frame $t$ in the $k$-th level of hierarchical Conv LSTM network and the output of the $k$-th ConvLSTM layer for frame $t$ as $y_{t,k}$.
As shown in Equation (1-3), $y_{t,k}$ depends on three variables: 
(1) the output of the previous ConvLSTM layer $y_{t,k-1}$;
(2) the extracted features in hierarchical ConvLSTM from the encoder $g_{t, k}$; and
(3) the hidden state of the same ConvLSTM layer for the previous frame $y_{t-1,k}$.

\begin{equation}
\centering
h_{input}=[g_{t, k}\ |\ y_{t,k-1}]
\end{equation}
\begin{equation}
\centering
h_{state}=y_{t-1,k}
\end{equation}
\begin{equation}
\centering
y_{t,k} = ConvLSTM_{k}(h_{input}, h_{state})
\end{equation}

In Equation (3), $h_{input}$ is the input of ConvLSTM, and $h_{state}$ is the hidden state input of ConvLSTM. 
$[A|B]$ is the concatenate operation for tensor $A$ and $B$ on feature axis.
For the first frame of a CMR sequence, $h_{state}$ is a matrix of ones, which means no prior information is known.

\begin{figure*}[]
  \centering
  \includegraphics[width=1.9 \columnwidth]{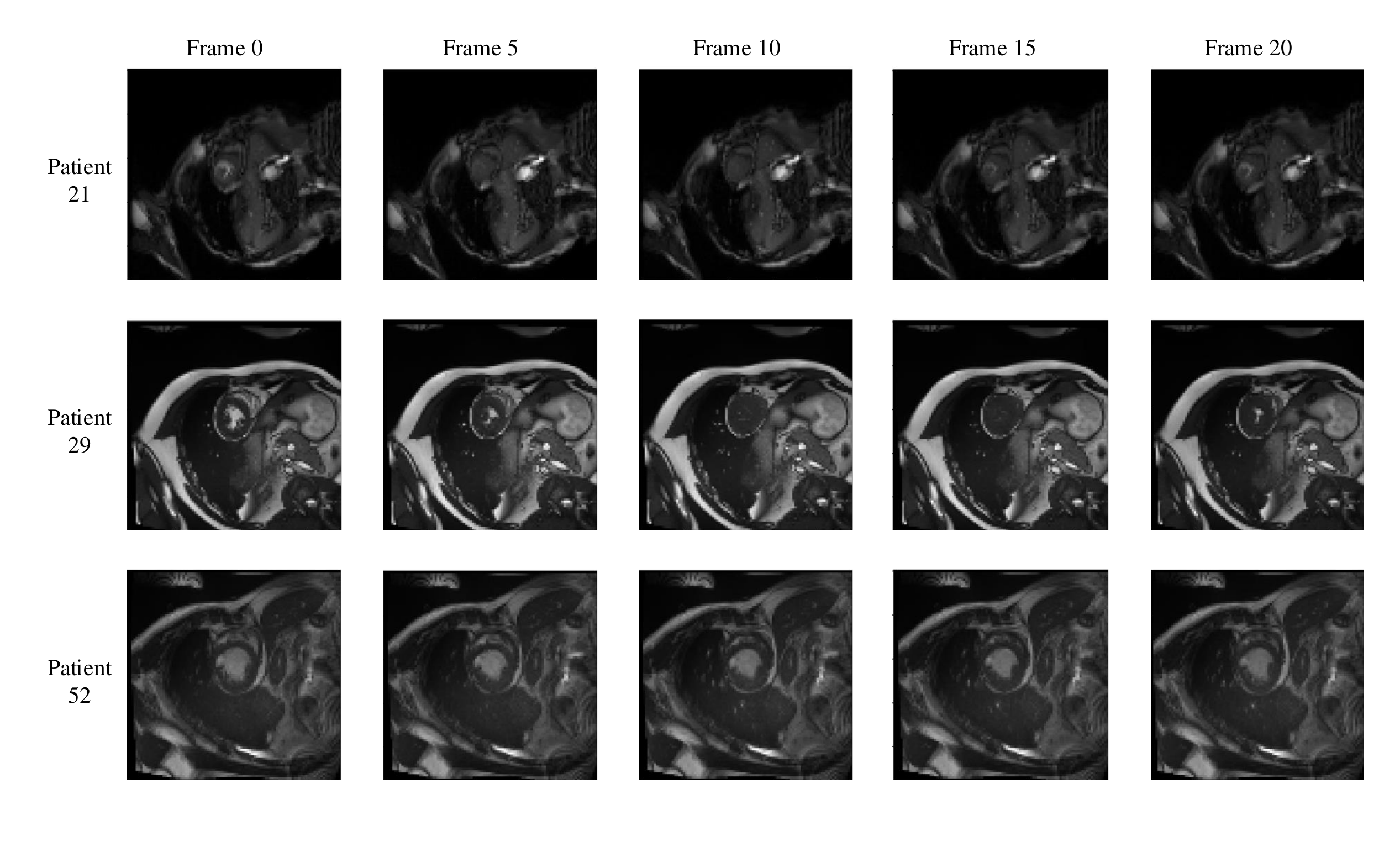}
  \caption{Frames of CMR sequence from three patients. Note the brightness heterogeneity in LV and RV on the first few frames. Using LSTM as decoder, the model can get more temporal information from the previous and future frames and result in more accurate segmentation.}
  \label{fig6}
\end{figure*}

\subsection{Bi-directional Training}

We notice that the prediction results of myocardium segmentation in CMR image is highly related to the segmentation result of frames either behind or after it. The first frame of CMR sequence will not receive enough information if we only use forward Conv LSTM.
Figure \ref{fig6} shows some frames of CMR image of different patients in the ACDC dataset.
We can see that the frames of CMR image of the last frame is highly related to the image of the next frame.
Consequently, the prediction error of the first frame due to the brightness heterogeneity may propagate to the rest of the CMR frames.

Therefore, we adopted a bi-directional training approach to alleviate this problem. 
Specifically, we used two Conv LSTMs in our decoder model. One will propagate forward, from frame 1 to frame $T$, while the other will propagate backward, from frame $T$ to frame 1.
$T$ is the total number of frames in one CMR sequence.
Figure \ref{fig7} presents the workflow of proposed bi-directional training approach.
Such approach can better exploit the temporal information along the frames thus improve the segmentation accuracy.

\section{Experiments}

\subsection{Experiment Setup}

In this section, we evaluated the performance of our proposed encoder-decoder framework in myocardial segmentation task of CMR sequences.
The residual U-net (Res U-net) implementation is used as our baseline. 
We compared the proposed framework (Res U-net+ConvLSTM) and Res U-net. 
Res U-net and ConvLSTM are implemented using PyTorch based on \cite{isensee2018nnu} and \cite{xingjian2015convolutional}, separately. 
The CMR images are resampled into $96\times 96\times 24$ using linear resample method. 
For data augmentation during training, we scaled all images by $0.8$ and $1.2$ and flipped them on $x$ axis and $y$ axis respectively. 
During testing and validation, we did not employ any augmentations. 
For each iteration, a complete CMR sequence containing 28-40 frames of a patient was used for training. 
Batch size is set to 1, which means in each iteration 1 CMR sequences containing 28 - 40 frames are fed for training.
We trained the encoder for 10 epochs with a learning rate of 0.0001. Then, we trained the encoder and decoder together with a learning rate of 0.0001 and a learning rate decay of 0.7 per epoch for another 10 epochs.

We splitted the ACDC dataset into training set, validation set and testing set by a ratio of 7:2:1 based on the patient number. 
Therefore, there are 70 patients in the training set, 20 patients in the validation set, and 10 patients in the testing set.
Dice coefficient and Intersection over Union (IoU) are used to evaluate the segmentation performance, which are defined as:

\begin{equation}
Dice(P,T) = 2\times \frac{\sum_{n=1}^{N}{(P_n\times T_n)}}{\sum_{n=1}^{N}{(P_n + T_n)}}
\end{equation}

\begin{equation}
IoU(P,T) = \frac{\sum_{n=1}^{N}{(P_n\times T_n)}}{\sum_{n=1}^{N}{(P_n + T_n-P_n\times T_n)}}
\end{equation}
in which $P$ and $T$ refer to the prediction and ground truth mask, respectively. $n$ is the index of all voxels (totally $N$ voxels). 

\begin{figure*}[]
  \centering
  \includegraphics[width=2\columnwidth]{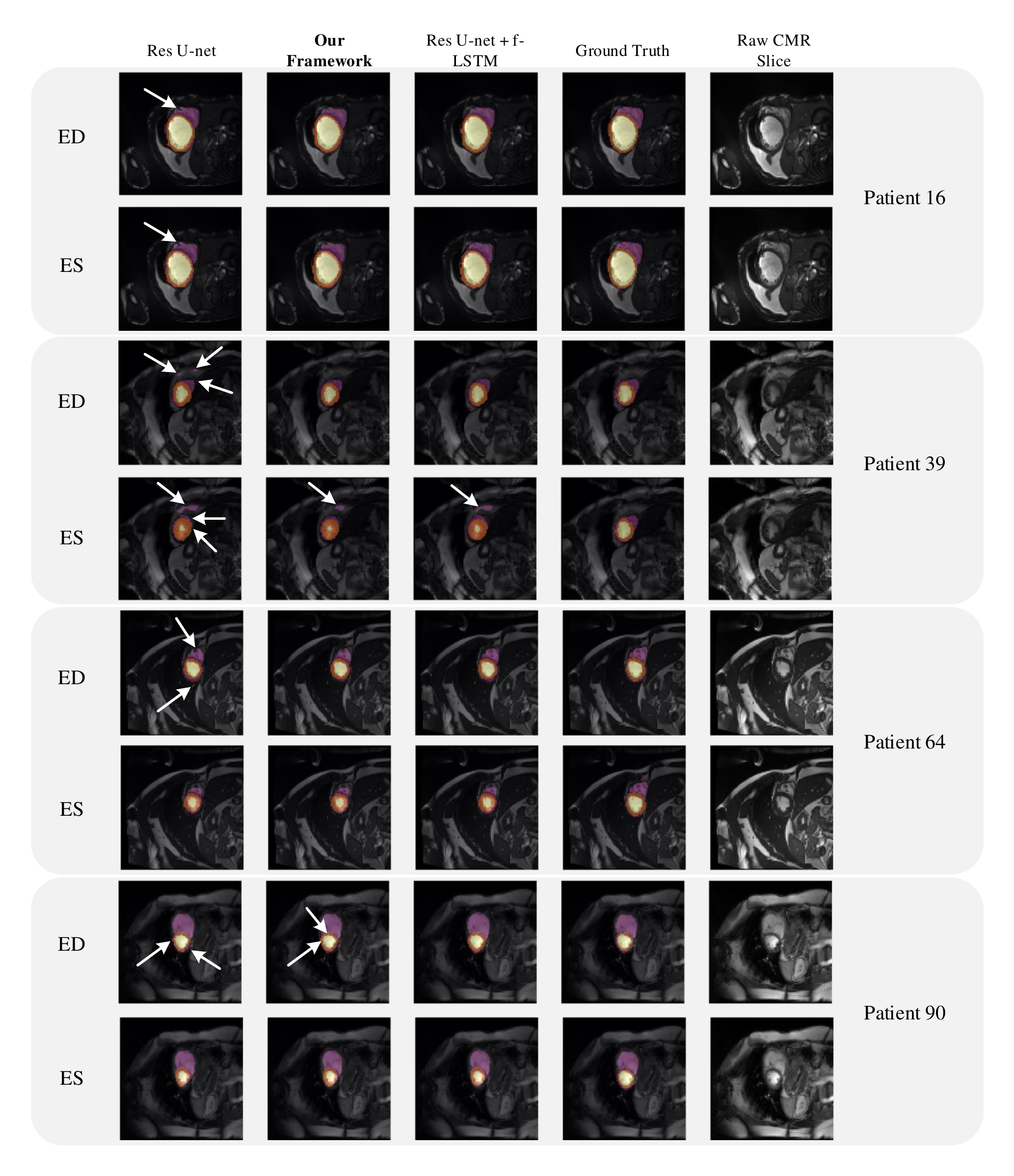}
  \caption{Visualization of CMR image segmentation results of three different patients in both ED phase and ES phase. Yellow, orange, and purple areas refer to the LV, MYO, and RV respectively. Each row refer to segmentation result of Res U-net, our framework, Res U-net + f-ConvLSTM, and ground truth from left to right. The white pointers in the image specifically point out the segmentation result that is inconsistent. From the figure we can see the f-ConvLSTM and bi-ConvLSTM model, which incorporate the temporal information between frames can greatly decrease the existence of such inconsistent segmentation result. Also, most errors in Res U-net + fConvLSTM and our framework are in hard cases like Patient 39 in this figure, where input CMR image has a low contrast and vague contour between labeled tissue and background tissue.}
  \label{fig8}
\end{figure*}

\begin{table*}[]
\centering
\caption{
  Comparison of our proposed framework against Res U-net baseline implementation. Res U-net+f-ConvLSTM refers to have only forward ConvLSTM that trains forwardly from frame 1 to frame $T$. Res U-net+bi-ConvLSTM (our framework) refers to training both forward ConvLSTM and backward ConvLSTM, where backward ConvLSTM process input CMR image from frame $T$ to frame $1$.
  }
\resizebox{\textwidth}{!}{
\begin{tabular}{llllllllll}
\hline
  & \multicolumn{4}{c}{\textbf{ED}} & \multicolumn{4}{c}{\textbf{ES}}\\
  & LV-Dice & RV-Dice & MYO-Dice & IoU & LV-Dice & RV-Dice & MYO-Dice & IoU \\
\hline
  Res U-net\cite{ronneberger2015u} &                   0.8856  & 0.8073  & 0.7178 & 0.5583 & 0.8050 & 0.6841 & 0.7554 & 0.4053\\
  Res U-net\cite{ronneberger2015u} + f-ConvLSTM\cite{salvador2017recurrent} &      0.8857  & 0.8082  & 0.7097 & 0.5586 & 0.8056 & 0.6896 & 0.7588 & 0.4186\\
  our framework &     \textbf{0.8967} & \textbf{0.8146} & \textbf{0.7260} & \textbf{0.5587} & \textbf{0.8133} & \textbf{0.7080} & \textbf{0.7656} & \textbf{0.4231}\\
\hline
\end{tabular}}
\end{table*}

\begin{table*}[h]
  \centering
  \caption{Quantitative segmentation results of different models for some frames corresponding to Figure \ref{fig8}.}
  \normalsize{
  \begin{tabular}{llllllll}
  \hline
      & & \multicolumn{2}{c}{\textbf{Res U-net}}   & \multicolumn{2}{c}{\textbf{Res U-net+f-ConvLSTM}}   & \multicolumn{2}{c}{\textbf{our framework}}  \\
      & & IoU & Dice   &  IoU & Dice  &  IoU & Dice \\
\hline
  \multirow{2}{*}{Patient 16} 
  & ED & 0.5417 & 0.8245 & 0.5417 & 0.8433 & 0.5417 & 0.8499 \\
  & ES & 0.5729 & 0.8168 & 0.5729 & 0.8196 & 0.5729 & 0.8234 \\
  \multirow{2}{*}{Patient 39} 
  & ED & 0.6771 & 0.8132 & 0.6771 & 0.8364 & 0.6771 & 0.8348 \\
  & ES & 0.7396 & 0.7556 & 0.7396 & 0.7688 & 0.7396 & 0.7587 \\
  \multirow{2}{*}{Patient 64} 
  & ED & 0.6666 & 0.8537 & 0.6667 & 0.8653 & 0.6667 & 0.8616 \\
  & ES & 0.6875 & 0.8334 & 0.6875 & 0.8368 & 0.6875 & 0.8347 \\
  \multirow{2}{*}{Patient 90}
  & ED & 0.6563 & 0.8099 & 0.6563 & 0.8027 & 0.6563 & 0.8043 \\
  & ES & 0.6563 & 0.7476 & 0.6563 & 0.7482 & 0.6563 & 0.7623 \\
  
  \hline
  \end{tabular}}
\end{table*}

\begin{figure*}[]
  \centering
  \includegraphics[width=2\columnwidth]{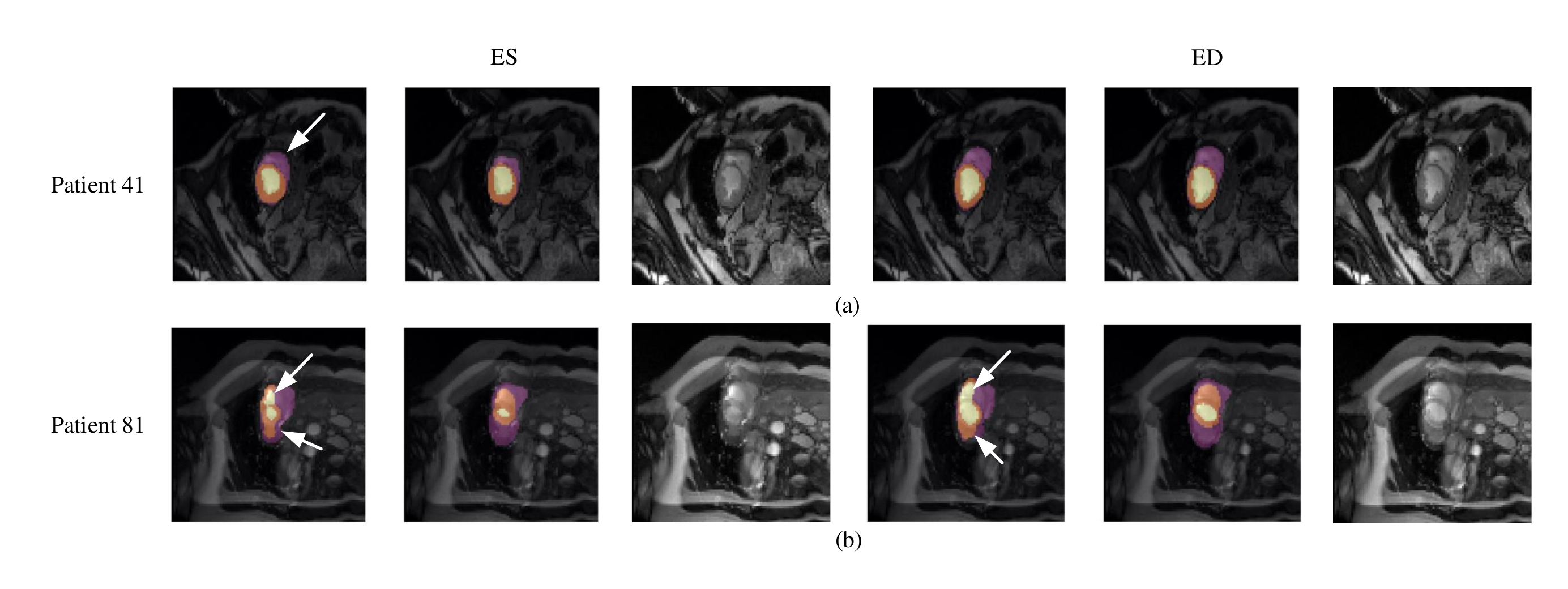}
  \caption{Typical segmentation error of our framework. The figure shows segmentation result for patient 41 and 81 in both ES and ED phases. The image on Column 1 and 4 are the segmentation results, and images on Column 2 and 5 are the corresponding ground truth. The segmentation error is usually caused by brightness heterogeneity, lack of contrast, or the improper input image due to faulty setup of magnetic resonance system or the misoperations of operators.}
  \label{fig9}
\end{figure*}

\subsection{Results and Discussion}

Table 1 shows the results of myocardial segmentation for LV, RV, and MYO at end diastolic (ED) phase and end systolic (ES) phase. 
Dice coefficient on each label class and Intersection over Union are reported.
Res U-net+f-ConvLSTM refers to training the proposed framework forwardly from frame 0 to frame $T$ while our framework (Res U-net+bi-ConvLSTM) refers to training in bi-direction.
We noticed that our framework outperforms baseline implementation in all metrics for both ED and ES frame.
Specifically, our Res U-net+f-ConvLSTM implementation has an improvement of 0.01\%, 0.10\%, -0.81\%, 0.61\%, 0.55\%, and 0.30\% on Dice coefficient of LV, RV, and MYO at ED frame and ES frame respectively.
The experiment results of Res U-net+f-ConvLSTM and our framework show that by adding a backward training step, we can further increase the segmentation performance.
Our framework's implementation has an improvement of 1.11\%, 0.64\%, 0.82\%, 0.83\%, 2.39\%, and 1.02\% on Dice coefficient of LV, RV, and MYO at ED frame and ES frame respectively.

Figure \ref{fig8} shows the visualization of segmentation results of four different patients (Patient 16, Patient 39, Patient 64, and Patient 90) in both ED phase and ES phase by Res U-net, our framework, and Res U-net+f-ConvLSTM.
Each row refers to the result of Res U-net, our framework, Res U-net+f-ConvLSTM, ground truth, and raw CMR slice respectively.
From Figure \ref{fig8}, we can see the f-ConvLSTM and bi-ConvLSTM, which has temporal consistency between frames, have less inconsistent segmentation result as marked by the white arrow in the figure. 
However, for a few cases, the temporal consistency may not eliminate the inconsistency in segmentation result completely. This happens when the framework recognize a stable feature on CMR image as an incorrect label. Since the misleading structure will remain on all the CMR frames in the sequence, the temporal consistency provided by LSTM will not be able to remove such an inconsistency.
Table 2 demonstrates the quantitative results for ED phase and ES phase of patient 16, 39, 64, and 90 corresponding to the Figure \ref{fig8}.
We can see that our proposed encoder-decoder framework tends to predict more consistent and accurate than the baseline, especially in the first few frames such as ES phase of patient 16, where obvious defect exists in segmentation result.
Comparing Res U-net+f-ConvLSTM and our framework on segmentation boundaries, we can observe that the bidirectional training can help our framework to produce more consistent results in most cases.
Although in some cases the segmentation of Res U-net+f-ConvLSTM is better than that of our framework, this might be caused by the constraint from backward temporal information that makes the segmentation lack flexibility.
The overall performance of our framework is superior in terms of all the metrics.

It can be seen that the Dice coefficient of the ED phase are usually higher than the ES phase.
However, our framework can achieve higher performance on both phases compared with Res U-net implementation.

Note that the work \cite{Yang2018} used the class-balanced loss and transfer learning to improve the performance of Res 3D U-net on the ACDC dataset. 
They achieved Dice coefficients of 0.864, 0.789, 0.775, and 0.770 on segmentation of LV and RV in ED phase and ES phase respectively, while our framework achieves higher Dice coefficients of 0.897, 0.815, 0.813, and 0.708, respectively.


\subsection{Discussion}

Quantitative segmentation results and grading results demonstrate the superiority of our framework compared with the Res U-net baseline implementation.
However, there are still some cases where our framework cannot predict reasonable boundaries. For example, Figure \ref{fig9} (a) shows the segmentation results of Patient 41 in ES and ED phases by our framework.
We can notice that there exists deviation between the ground truth boundary (images on Column 2 and 4) and the prediction boundary (images on Column 1 and 3).
This is because the flow of blood in RV cavity leads to the brightness heterogeneity in the RV area of CMR image, which makes the image intensity of the ground truth RV region similar to the surrounding cardiac structures (e.g., heart chambers), and finally leads to segmentation failure.

There are some cases as shown in Figure \ref{fig9} (b) in which CMR  sequence have serious defect and ghosting.
This may be caused by the improper setup of the magnetic resonance system or the mistake of operators.
Therefore, it is hard for our framework to find a plausible myocardial boundary even though our framework is able to correct some segmentation error based on temporal information between frames.

\section{Conclusion}

In this paper, we proposed a myocardial segmentation framework of CMR sequences for CAD diagnosis. 
Specifically, we proposed to combine conventional networks and recurrent networks to incorporate temporal information between sequences to ensure temporal consistency.
Extensive experiments showed that compared with Res U-net, our proposed framework can achieve an improvement of 1\% to 3\% in Dice coefficient.
In addition, we introduced a bi-directional training approach to further reduce segmentation error introduced by the first few frames in the forward training process.
Experiment results demonstrate that our bi-directional training approach can further improve the segmentation performance.



{\small
\bibliographystyle{ieee_fullname}
\bibliography{main.bbl}
}

\end{document}